\documentclass[%
 aip,
 amsmath,amssymb,
 reprint,%
]{revtex4-1}

\usepackage{graphicx}
\usepackage{dcolumn}
\usepackage{bm}

\usepackage[utf8]{inputenc}
\usepackage[T1]{fontenc}
\usepackage{mathptmx}
\usepackage{amsmath}
\usepackage{physics}
\begin{document}

\preprint{AIP/123-QED}

\title[Ordinal Patterns in the Duffing Oscillator]{Ordinal Patterns in the Duffing Oscillator:\\Analyzing Powers of Characterization}

\author{I. Gunther}
 \email{ivanawg@gmail.com}
\author{Arjendu K. Pattanayak}%
\email{arjendu@carleton.edu}
\affiliation{ 
Department of Physics and Astronomy, Carleton College, 1 N College St, Northfield, MN 55057 
}%

\author{Andr\'es Aragoneses}
\email{aaragoneses@ewu.edu}
\affiliation{%
Department of Physics, Eastern Washington University, Cheney, WA 99004 
}%

\date{\today}

\begin{abstract}
Ordinal Patterns are a time-series data analysis tool used as a preliminary step to construct the Permutation Entropy which itself allows the same characterization of dynamics as chaotic or regular as more theoretical constructs such as the Lyapunov exponent. However ordinal patterns store strictly more information than Permutation Entropy or Lyapunov exponents. We present results working with the Duffing oscillator showing that ordinal patterns reflect changes in dynamical symmetry invisible to other measures, even Permutation Entropy. We find that these changes in symmetry at given parameter values are correlated with a change in stability at {\em neighboring} parameters which suggests a novel predictive capability for this analysis technique. 
\end{abstract}

\maketitle

\begin{quotation}
Nonlinear dynamical systems can be regular, stable, and simple as well as chaotic, unstable, and complex, sometimes switching between the two behaviors abruptly as a function of parameters. Among the techniques useful in distinguishing between these very different dynamics is computing the so-called permutation entropy (PE) for the system. Here the time-series associated with a dynamical observable is first converted into a discrete set of observations about extreme events (maxima or minima, for example), and the relative length of the time-spacing between these events is then used to construct `ordinal patterns' or Words. The PE is the entropy associated with the probability distribution of these Words. The probability distributions themselves contain strictly more information than does the PE. In our work we show that the probability distributions of Words can be analyzed to discover changes in dynamical symmetry in chaotic regimes (and stable regimes). We show that in fact the growth of symmetry in Word probabilities are quantifiably more frequent when the system is about to transition to stable behavior at a nearby parameter, thus allowing for the possibility of predictions beyond that of the usual techniques.
\end{quotation}

\section{\label{sec:level1}Introduction}
Nonlinear dynamical systems display a rich variety of behaviors. Within such systems the dynamics can be regular, repeating periodically where the periodicity can be simple or high-order, or else repeating quasi-periodically. These dynamics can also be stable, such that a small perturbation or error in specification of initial conditions do not grow or qualitatively change the behavior. Long-term predictability is only possible under these circumstances. However, such systems often evolve in a chaotic (bounded and non-periodic) and unpredictable manner such that small errors do grow exponentially rapidly; this is quantified by the existence of a positive definite maximal Lyapunov exponent (LE) of the dynamics which measures the sensitivity to initial conditions and is the inverse time-scale for predictability. Chaotic trajectories are algorithmically complex \cite{Boffetta(2002)} in that any recorded time-series for a given observable cannot be compressed by an arbitrary factor. In contrast, regular systems do not have positive LEs, and are algorithmically simple. Typically we see both kinds of behavior: If there are co-existing attractors, this can occur as a function of initial conditions. In other situations the dynamics can switch back and forth between the two abruptly as a function of system parameters. The sensitivity to change of parameters can also strongly affect controllability and dynamical response \cite{Shinbrot(1993), Hayes(1993), Cavalcante(2013), Wang(2016)}.

Since such nonlinear systems are ubiquitous in nature, occurring for example in mechanical structures such as bridges, and in physiological systems, apart from the wide range of physics applications, identifying the difference between the two kinds of behavior and the possibility of a transition between them can therefore be of literally vital importance. However, LEs are difficult to access in experimental or observational situations since they are a theoretical construct that requires on access to the fiducial dynamical trajectory itself {\em and} its tangent space -- that is, its response to infinitesimally small perturbations. Techniques based on extracting the informational complexity directly from experimental observations have been developed over the years to address this challenge \cite{Wolf(1985), Bandt(2002), Amigo(2006), Amigo(2010), Zou(2019)}. In particular, the Permutation Entropy (PE) \cite{Bandt(2002)} computed from the probability distribution of ordinal patterns (or `Words', defined more carefully below) constructed from the time series has been shown to be a suitable alternative for LEs, particularly in following parameter dependencies, including applications to semi-classical problems \cite{Trostel(2018)}. Intriguingly, in constructing the PE from ordinal patterns, a certain amount of information about the dynamics is compressed: That is, the probability distributions and particularly their dependence on parameters contain more information about the dynamics than the PE, such that it is possible to go beyond the simple classification into `chaotic' and `regular'. This suggests that ordinal patterns can be helpful in understanding the system's dynamics in situations involving experimental time-series of known dynamics and changeable parameters \cite{Soriano(2011), Barreiro(2011), Parlitz(2012), Tirabassi(2016), Colet(2018)}, where one might seek to understand the structured behavior of the system beyond the existence or lack of chaos, or where one might not have access to LEs.

In this paper we present results from a study of the complex dynamics of the paradigmatic Duffing oscillator aimed at characterizing this extra information and uncover some potential value of this information. We start from the observation that changes in Word frequencies arise from a change in temporal symmetries that are captured with some novel characterizations of Word probability distributions that use symmetry criteria. The behavior of these criteria as a function of parameters is visually observed to be -- and then quantifiably correlated with -- a chaos-stability transition at {\em neighboring} parameters. When we compare the results with some well known techniques (LEs, PE, Poincar\'{e} Sections, Fourier spectra of time-series, and phase-space diagrams) we do not find this correlation in these other measures. We then dig further into approaches intended to capture temporal correlations in the time series to see if the patterns visible in Word populations can be understood better. Specifically, we show Word probabilities can be derived from Interval Trio Maps (ITM), a generalization of a return map where the probability distribution of consecutive inter-event intervals is mapped into a higher-dimensional space, such that changes in probabilities correspond to change in clustering in sectors of the ITM space. We also consider Generalized Poincar\'e Sections (GPS) which can be understood as the projection of the dynamics into a subspace following an event-driven criterion. While these latter new analysis techniques contain novel information complementary to that from other techniques, they provide similar results and also seem to fail to display the correlation seen in Word probability distributions. Thus, when analyzing the probability distribution of the ordinal patterns, we not only capture signatures of temporal correlations and memory in complex dynamical systems\cite{Barreiro(2011), Parlitz(2012), Colet(2018)} but also find that they extract information about the system not available with other techniques, in particular to provide signals about qualitative dynamical transitions. 

The paper is laid out as follows: We first describe the Duffing oscillator and the known analysis tools we use to study it, focusing in particular on the permutation entropy and ordinal patterns, and present our initial results from scanning parameter space. We further describe the detailed techniques for analysis of Word populations where we introduce new symmetry-based metrics for Word populations that help unveil hidden structure in the system. We then introduce the new analysis techniques -- the GPS, and ITM -- and compare their results to those from Words. We conclude with a short discussion including the potential applicability of these new techniques.

\section{Words in the Duffing Oscillator}
The Duffing oscillator is a paradigmatic model\cite{Guckenheimer(1983)} of a damped and periodically driven nonlinear oscillator, completely characterized by trajectories in a phase-space defined by position, momentum, and time, or $x[t]$, $p[t]=\dot{x}[t]$, and $t$. We study here the bi-stable version, where the oscillator can in principle travel between two potential wells. These dynamics are defined by
\begin{equation}
    \ddot{x}+d\dot{x}+ax+bx^3=g\cos{\omega t},
    \label{eq:Duf}
\end{equation}
where we have chosen default dimensionless constants of stiffness $a=1$, nonlinearity $b=-1$ and damping $d=0.3$. For the detailed studies we present below, we also fix the external driving frequency at $\omega=1.2$ and scan its amplitude $g$ across a range chosen to show a variety of behaviors. Generically, all results carry into other parameter regimes, even though we do not present them here. In this system's parameter space, there exist distinct regions or bands of chaos and regularity, where the regularity can arise from global stable attractors which are either simple or high-order periodic orbits. We take care to simulate the dynamics to discard initial transient behavior (though there always remain the possibility of `long transients') by using simulation runs of $t=10000$. This is particularly important, for example, in constructing Lyapunov exponents which we do using the method laid out by Wolf\cite{Wolf(1985)}. 

To construct Words or ordinal patterns, we record the time-values of local maxima or 'peaks' in $x(t)$, and track the lengths of time between peaks, or 'interpeak intervals'. Sequences of interpeak intervals determine the Words. That is, Words are defined by considering groups of $n$ consecutive or semi-consecutive (where one would skip every other, or every third) events in a time-series, which are then ranked within the group according to some selected criteria\cite{Bandt(2002)}. In our case, we choose $n=3$ so that the Words are 3-dimensional (i.e., they are constructed from trios of events), consecutive (not skipping regular events), and rank interpeak intervals. For instance, if the first of three interpeak intervals is of middling length, the second is the shortest of the three, and the third is the longest, this would be categorized as a "1-0-2" type Word, so that we have one incident of the the word $w_{102}$. It is easy to see that there are thus six types of 3D Words: $w_{012}$, $w_{021}$, $w_{102}$, $w_{120}$, $w_{201}$, and $w_{210}$, whose relative populations across the time-series is the quantity of interest. We choose $n=3$ for practical reasons: We can confirm empirically that 2-dimensional Words do not yield detailed structure, there being only two types, whereas 4-dimensional and beyond subdivide dynamics too much to be useful. We should also note that marginal cases where two or more intervals in the trio were of equal length within the accuracy of our simulation are recorded as being $w_{111}$ or periodic Words. In what follows, we have excluded periodic Words, though depending on the goal of the Words analysis, some researchers choose to remove them, add them to another Word's population, or treat them as experimental uncertainty; we also note that in our case they are well-defined and abundant. Already, Words have been fruitfully applied to the Duffing oscillator to show that Permutation Entropy, derived from Word populations, tracks chaos and stability much like Lyapunov exponents\cite{Bandt(2002)}, and detects certain changes in the dynamics that LE cannot\cite{Trostel(2018)}. We note that the attempt to find further structure in Word populations beyond the PE is not new: For example, Bandt\cite{Bandt(2015),Bandt(2019)} has studied spatial 3D Words, and constructed various autocorrelation functions from their populations, discussed below.

\begin{figure}
    \centering
    \includegraphics[width=250pt]{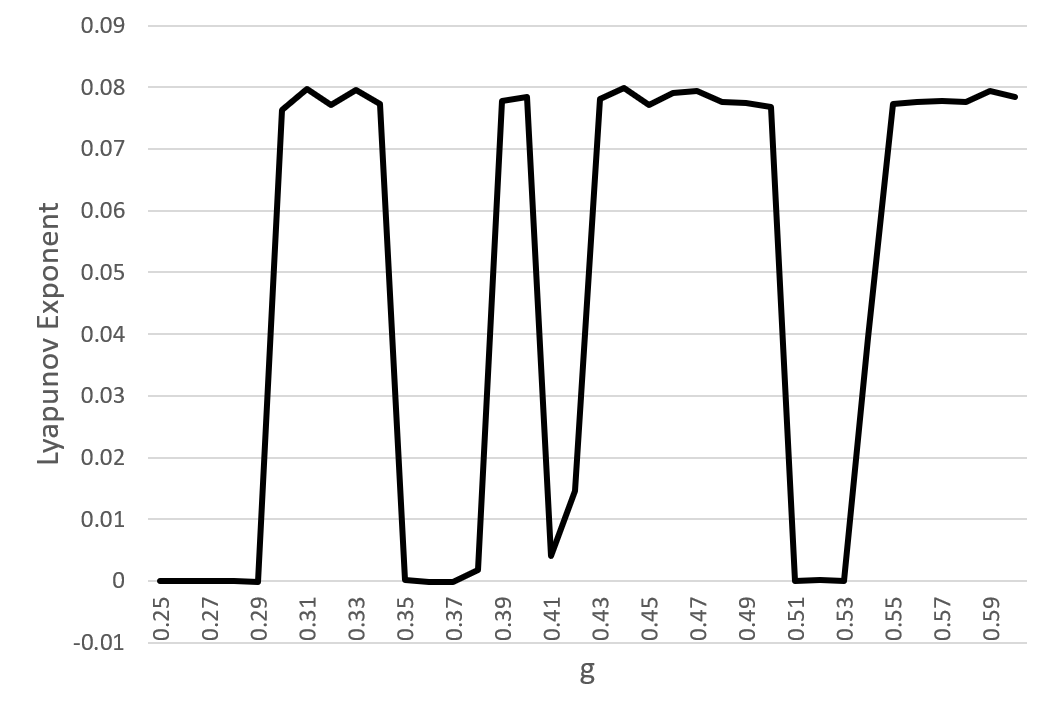}
    \caption{The $g$-range of the Lyapunov exponent. At the $g$-values of $0.41$, $0.42$, and $0.54$, the chaos is transient, as confirmed by running the oscillator slightly beyond $t=10000$.}
    \label{fig:LEg}
    \includegraphics[width=250pt]{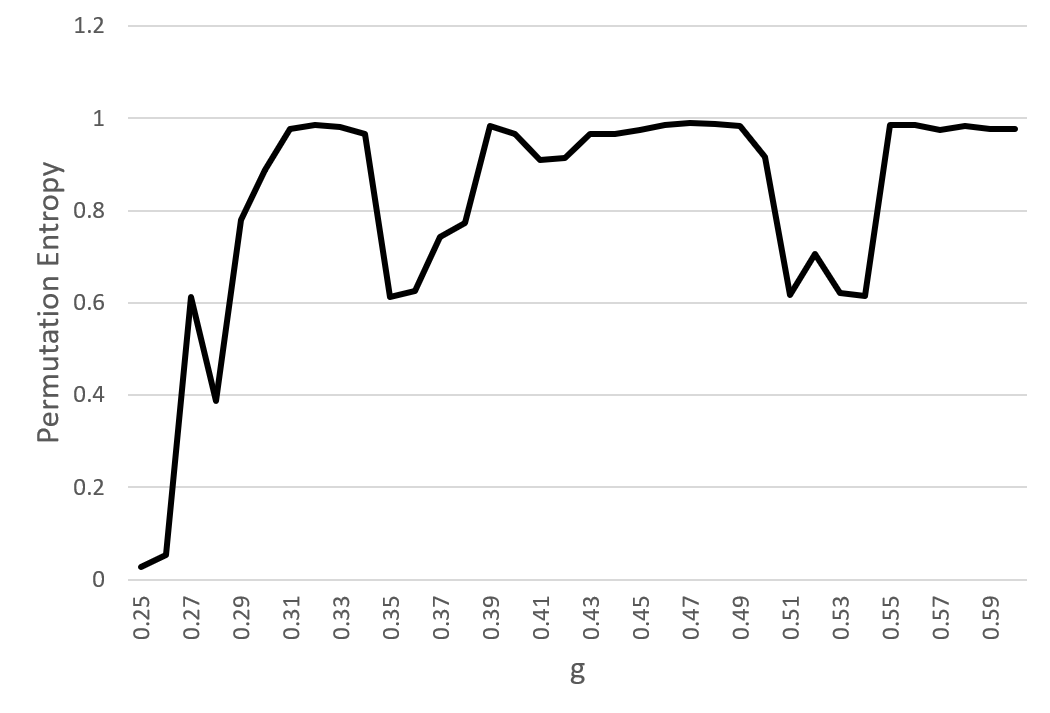}
    \caption{The $g$-range of Permutation Entropy, calculated from Word populations. Note the broad agreement with LE, in that parameters with low LE tend to have Permutation Entropy below $0.9$.}
    \label{fig:PEg}
\end{figure}

\begin{figure}
    \centering
    \includegraphics[width=250pt]{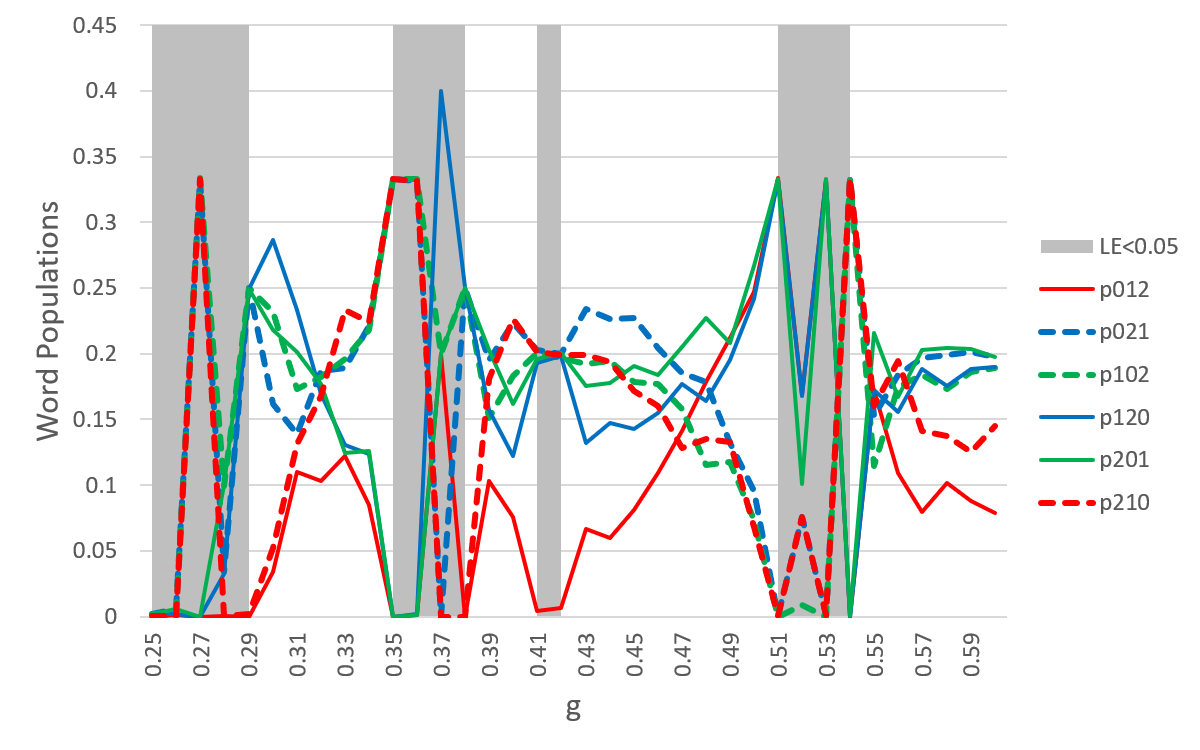}
    \caption{The $g$-range of word frequencies, with stable or transiently chaotic parameters highlighted in grey. Note that in some cases, the sum of the populations is less than 1, due to an abundance of consecutive interpeak intervals that were nearly exactly equal ($w_{111}$ not plotted), and therefore could not be ordinalized. The $g$-values $0.30$, $0.34$, and $0.50$ are chaotic, as shown in Fig. \ref{fig:LEg}, but exhibit certain groupings that mimic their stable neighbors.}
    \label{fig:wordg}
\end{figure}

We start with the initial results from our scan of parameters, focusing on the varying $g$ from $0.25$-$0.60$ in increments of $0.01$; this range shows several transitions between bands of regular to chaotic behavior of relevance to our study. We show here LEs (see Fig.\ref{fig:LEg}), Word populations (see Fig.\ref{fig:wordg}), and PE (see Fig.\ref{fig:PEg}) which is calculated from Word populations $P_{w_i}$ for each word $w_i$ of dimension $D$ as 
\begin{equation}
    PE = -\sum_{i}^{D!}\frac{P_{w_i} \log P_{w_i}}{\log (D!)}.
\end{equation}
This construction makes clear how the PE embodies a compression of information about Word populations, whence there is arguably more information about the dynamics in the distribution of Words than is visible in the PE. We also note that to better understand our results, we have also used other diagnostics: Phase-space trajectory diagrams (not shown), Poincar\'{e} Sections (PS), Fourier Transforms (FT) (not shown), for example, but here we start with a summary and comparison of Word populations with the traditional measure of LEs. These exponents measure the sensitivity to perturbation of the dynamics, and characterize the (exponential) rate which two slightly different initial conditions separate in phase-space. While the number of Lyapunov exponents depends on the number of dimensions in the dynamics, the focus is typically on the largest, which is positive for a chaotic system. We find indeed that LE and PE scans show the same structure of chaos and stability bands, although their agreement is imperfect in the presence of long transient chaos. We term this behavior as long transient chaos because the dynamics collapse into stability sometime {\em after} $t=10000$ time, unlike stable regions which manifest as regular much faster, and fully chaotic regions which remain so for seemingly indefinite times. Long transient chaos at $g=0.41-0.42$ is nearly indistinguishable from chaos through the PE, but at $g=0.54$, the PE treats it as full stability, even though the corresponding LE is higher than at $0.42$. 

The scan of Word populations, on the other hand, stand out from the PE and LE scans. Not only does this follow the transitions from regularity, it also points to the existence of {\em different} types of chaos (as well as {\em different} types of stability). For example, compare $g=0.32$ with $g=0.34$, which are both chaotic according to both the LE and the PE calculations. However, at $g=0.32$ we find that the Word populations are distributed roughly evenly such that their values could be the result of random chance, while at $g=0.34$ the dynamics shows a clear systemic preference for certain three Words over certain other three, a preference that becomes even more clear and absolute as the dynamics become stable at $g=0.35$. In fact this pattern persists in general: Word populations seem to become more ordered or structured often in the chaotic regime. Our first observation is that this increase in order seems to happen at parameter values which are near $g$ values where stable orbits exist (in this case at $g=0.35$) which is closer to $g=0.34$ than the more disordered $g=0.32$. This and other examples are visible in Fig. \ref{fig:wordg}.


\begin{figure}
\centering
  \includegraphics[width=120pt]{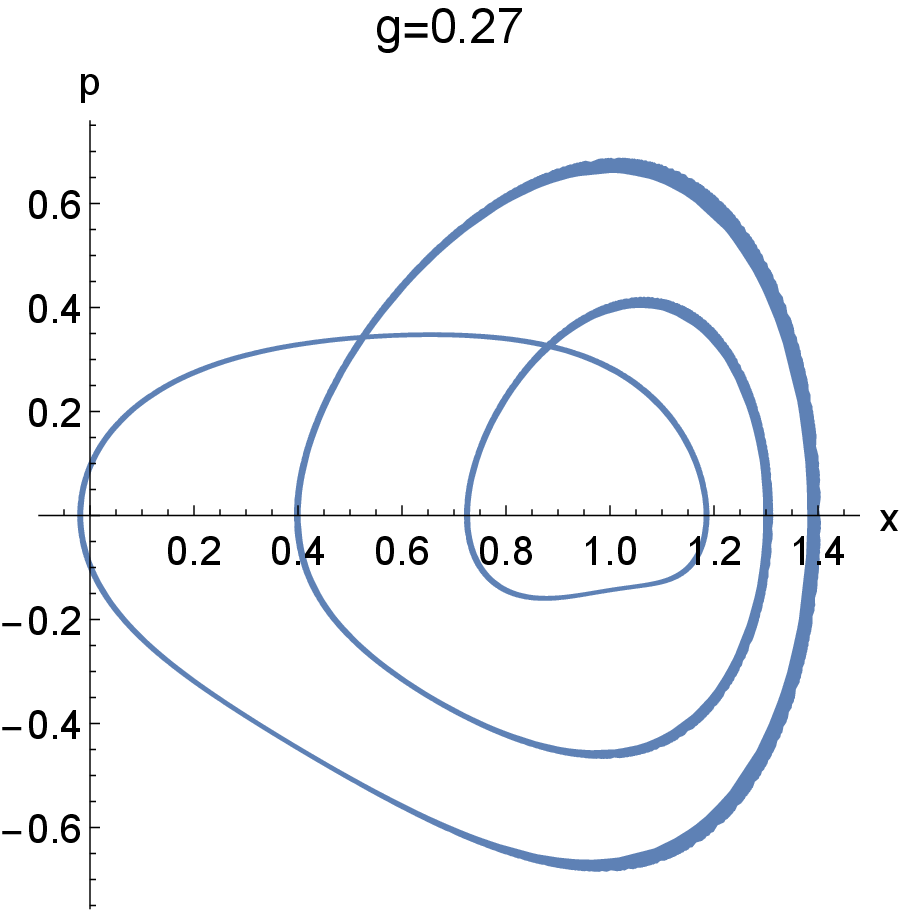}
  \includegraphics[width=120pt]{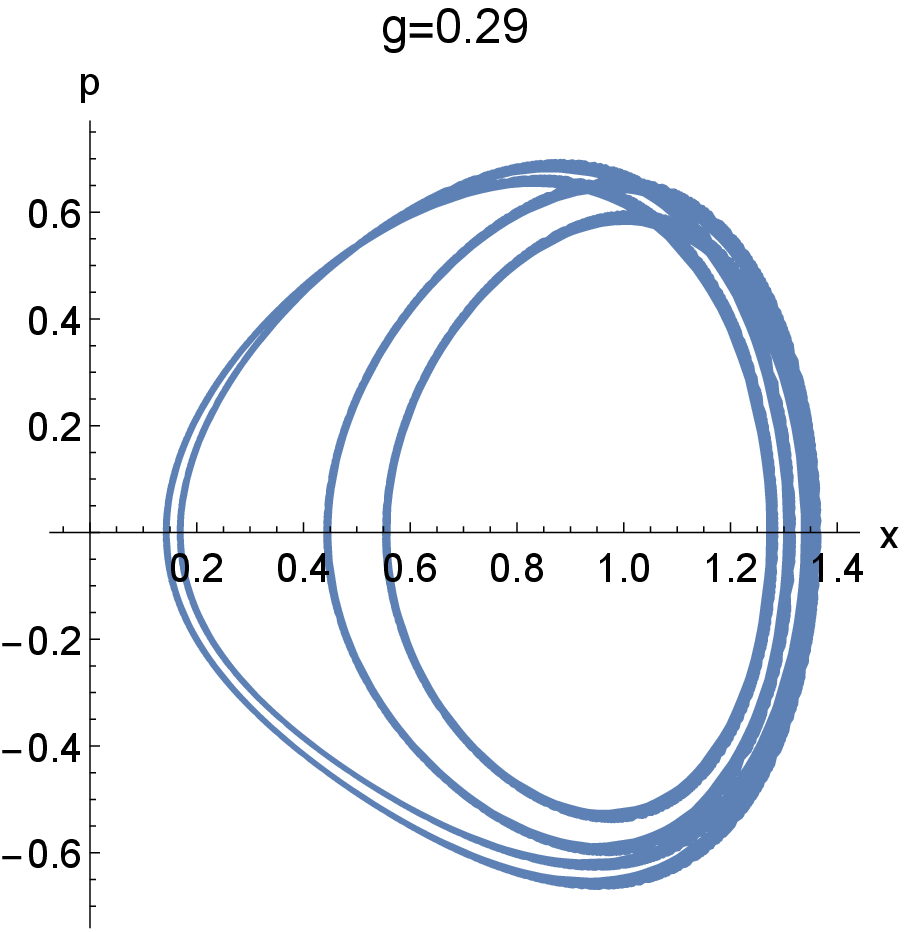}
\caption{The phase diagrams for $g=0.27$ (left) and $g=0.29$ (right), showing the physical phenomena which lead to Rotational and Mirror Symmetry, respectively. Note the three x-axis crossings at $g=0.27$, signifying a period-3 orbit, and the rough vertical symmetry of $g=0.29$, showing geometrical, if not topological, time-reversibility.}
\label{fig:pd}
\end{figure}

These results and in particular the change in the structure is still difficult to parse properly. To better clarify and understand this change in structure that Word populations exhibit in certain circumstances, we dig deeper using two methods of grouping Word populations using `symmetry groups' which we rigorously quantify and explicate in the next section, but motivate here. We have chosen these symmetries, as opposed to other inter-Word relationships, essentially empirically, based on the their apparent prevalence in the $g$-range (Fig. \ref{fig:wordg}), and their intuitive definitions. The first type of symmetry we consider is {\it{rotational symmetry}}, which occurs when the populations of $w_{012}$, $w_{120}$, and $w_{201}$ are similar to each other, as are those of $w_{021}$, $w_{210}$, and $w_{102}$, but these two population groups are dissimilar to each other. Mathematically, high rotational symmetry is characterized by
\begin{equation}
    P_{012}\simeq P_{120}\simeq P_{201}\neq P_{021}\simeq P_{210}\simeq P_{102}.
\end{equation}
This rotational symmetry is intuitively a {\em weak} indication of period-3 orbits in the time-series; that is, if each Word leads into another in a chain, e.g., $012012012$..., we would obtain an equal abundance of $w_{012}$, $w_{120}$, and $w_{201}$. Since Words are ordinal (that is, constructed from a coarse-graining of the time-series), and since the presence of these symmetries is relative but not absolute, it is not an absolute indicator of true period-3 orbits. The second type, {\it{mirror symmetry}}, occurs when populations of $w_{012}$ and $w_{210}$ are similar, as are those of $w_{021}$ and $w_{120}$, as are those of $w_{102}$ and $w_{201}$, and all three of these pairs are dissimilar to each other. Mathematically, high mirror symmetry would follow from
\begin{equation}
    P_{012}\simeq P_{210}\neq P_{021}\simeq P_{120}\neq P_{102}\simeq P_{201},
\end{equation}
corresponding to a tendency for small-scale events in the time-series to resemble themselves regardless of the direction of time. This is a weak form of time-reversal invariance. For example, we see that at $g=0.27$ and $g=0.35-0.36$, both of which appear only as stable to the LE (Fig. \ref{fig:LEg}), exhibit near-perfect rotational symmetry, wholly different from nearby $g=0.29$ and $g=0.38$, which are also simply stable according to the LE, but in contrast have equal populations of the two mirror symmetry groups, and total absence of the third. This symmetry is correlated to differences in the time-series and phase diagrams in the stable regime (Fig.\ref{fig:pd}). However, similar changes in symmetry do not visibly hold for chaotic phase diagrams.
Before we turn to the more detailed discussion of these symmetries in the dynamics, we remark that some types of stability prevent the formation of certain types of Words altogether: if the oscillator has perfect period-1 or period-2 orbits, then equality between nearby interpeak intervals will prevent ordinalization of 3D Words. This phenomenon is obvious in Word population graphics like Fig. \ref{fig:wordg}, because the sum of the six Words is significantly less than 1 at these locations, such as at $g=0.51$. This is because during the Word-counting process, we designated all marginal cases as "periodic Words" or $w_{111}$, which counts against the fractional populations of all other Words, but which is not shown in Fig. \ref{fig:wordg}.

\section{Quantifying Symmetry}
We can formalize the notions of the existence of rotational and mirror symmetry groups in the dynamics by computing statistical variances
\begin{equation}
    \sigma=\sum_i \frac{(P_i-\overline{P})^2}{6},
\end{equation}
which when applied to symmetry, give us four measures: Rotational Variance ($\psi$), Rotational Hierarchy ($\Psi$), Mirror Variance ($\xi$), and Mirror Hierarchy ($\Xi$). Rotational Variance tracks the variance of each word relative to the average of its Rotational Symmetry group, and Mirror Variance does the same with Mirror Symmetry groups. Likewise, Rotational Hierarchy measures the variance of each Rotational Symmetry group's average population, against the average for all six Words; similarly, Mirror Hierarchy measures each Mirror Symmetry group average. That is to say, if the population of certain Words agree with other Words in their symmetry group, then variance, $\psi$ or $\xi$, will be low. If the system has a clear preference for one Symmetry group over others, then the hierarchies, $\Psi$ or $\Xi$, will be high. The philosophy behind these symmetry groups is similar to that of Bandt's work \cite{Bandt(2015)} on autocorrelation functions defined as
\begin{eqnarray}
   &\beta=w_{012}-w_{210}\\
   &\tau=w_{012}+w_{210}-\frac{1}{3}\\
   &\gamma=w_{102}+w_{120}-w_{021}-w_{201}\\
   &\delta=w_{021}+w_{102}-w_{120}-w_{201},
\end{eqnarray}
where $\beta$ is termed Up-Down Balance, $\tau$ is Persistence, $\gamma$ is Rotational Asymmetry, and $\delta$ is Up-Down Scaling. There is a clear correspondence between the physical significance of Bandt's $\gamma$ and $\xi$, as time-reversal asymmetries, although Bandt names $\gamma$ as "rotational asymmetry". The advantage of this grouping and quantifying Words this way is that, if any meaningful underlying order in the Words increases near stability, it would be shown through these measures more clearly, and numerically, than by judging Word populations by eye. Thus, symmetry measures (specifically either $\Psi$ and $\psi$ or $\Xi$ and $\xi$) and Bandt's autocorrelation functions condense Word population information into a more concise, readable set of statistics, which are distinct from PE, and are useful in tracking temporal orderings and structure in the dynamics.

\begin{figure}
    \centering
    \includegraphics[width=250pt]{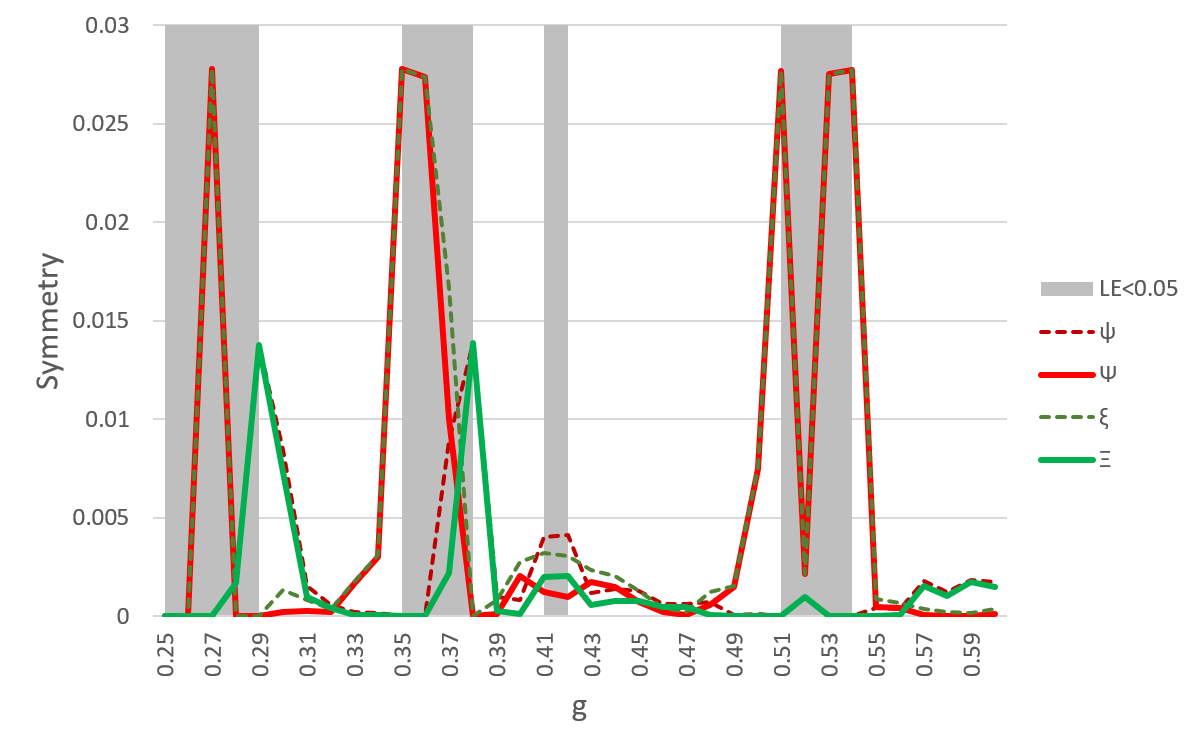}
    \caption{The $g$-range of Symmetry measures. Because of the orthogonality of Rotational and Mirror Symmetry groups, Rotational Variance often agrees with Mirror Hierarchy, likewise Rotational Hierarchy and Mirror Variance. However, each Variance is never less than the opposite Hierarchy.}
    \label{fig:sym0}
\end{figure}

\begin{figure}
    \centering
  \includegraphics[width=250pt]{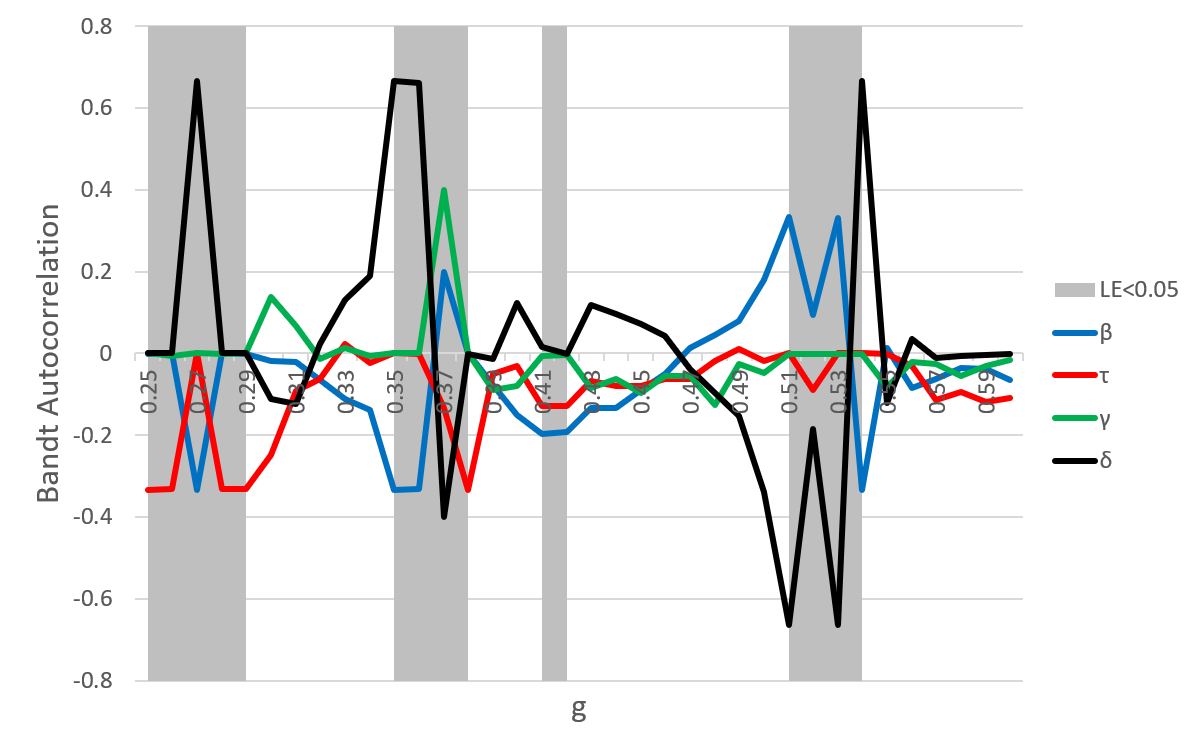}
    \caption{The $g$-range of Bandt autocorrelation functions, with stable g-values highlighted grey. In the Symmetries of Fig. \ref{fig:sym0}, note the increase in one Hierarchy, at $0.30$, $0.34$ and $0.50$, relative to their chaotic neighbors. Similarly in the autocorrelation functions, the absolute values of Up-Down measures $\beta$ and $\delta$ increase at the same points.}
    \label{fig:bandt}
\end{figure}

Figure~\ref{fig:sym0} quantifies these symmetries as a function of the control parameter $g$. In it we can appreciate how chaotic regions in which Words are somewhat ordered, such as $0.34$ and $0.50$, have significantly higher hierarchy and lower variance in one type of symmetry than in other chaotic regions (the latter therefore being arguably "truly" chaotic). Stable regions ($LE<0$) have even higher hierarchy. This brings home the significance of our result that -- at least on the basis of this exploration of the Duffing oscillator -- by measuring the symmetry of the dynamics we can conclude that the existence of a rotational or mirror hierarchy value between $~0.002-0.007$ or greater signifies the existence of stable parameter regimes in the neighborhood (thus, that this is `pre-stable chaos'), while hierarchy values closer to $0.01$ or greater represents true stability. In contrast, Fig.~\ref{fig:bandt} shows the autocorrelation functions introduced by Bandt in Ref.~\cite{Bandt(2015)}. They do not show to be suitable measures to discern the transitions present in this complex system. While it is true that there is an increase or decrease of $\beta$ or $\delta$ in the chaotic regime right before g=0.35 and g=0.51, there is no signature revealed right after g=0.29 or g=0.38.


\begin{figure}
    \centering
    \includegraphics[width=250pt]{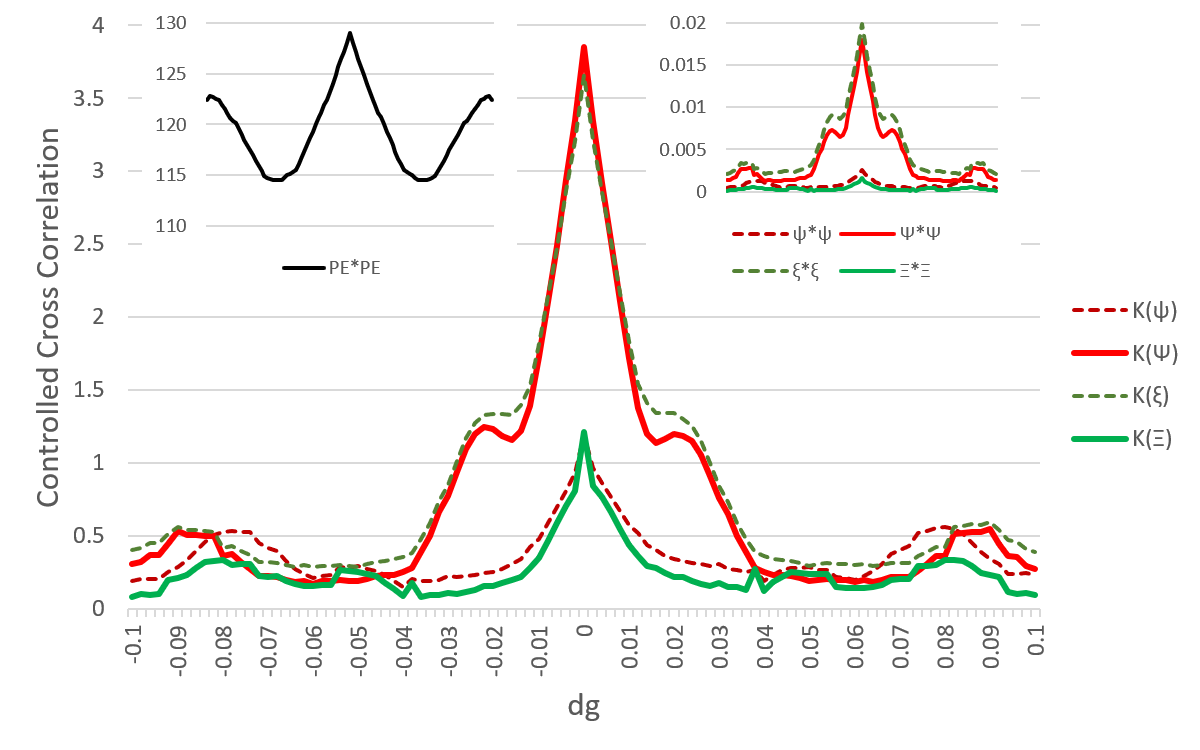}
    \caption{The Controlled Cross Correlation of the Permutation Entropy with the Symmetry measures, across offset $dg$, with insets of autocorrelations. The local minima around $\pm 0.015$ for $\Psi$ indicate that Rotational Symmetry increases near low Permutation Entropy with an offset of $g\approx 0.015$. This means that with some degree of accuracy, we can look around the corner for chaos-stability shifts using Symmetry.}
    \label{fig:ccc}
\end{figure}

It should be clarified that, as seen in Fig.\ref{fig:sym0}, this increase in order is markedly visible, but imperfect, as false negative and false positive signals of nearby stability occasionally occur. Explorations of the behavior of these quantities with variation in other parameters such as $\omega$ (not shown) also confirm this trend of Words always being ordered in the stable regime, as shown in Figs.\ref{fig:LEg} and \ref{fig:wordg}, but only being ordered within the chaotic regime typically {\em near} an upcoming transition to regularity. To go beyond this qualitative signal, we quantify the behavior as a measurable correlation between Symmetry and stability, as shown in Fig.~\ref{fig:ccc}.
To compute this, we first compute individual autocorrelations for the Permutation Entropy and the Symmetry measures $\psi$, $\Psi$, $\xi$, and $\Xi$, as functions of $g$ (Fig.\ref{fig:ccc}). We then compute the cross-correlation of the symmetry measures with the PE and construct a `controlled' Cross Correlation $K$ to control for the natural self-repetition scales in the Permutation Entropy and Symmetries so that finally the $K(X)$ for any Symmetry measure $X$ is
\begin{equation}
    K(X)=\frac{(PE\star PE)(X\star X)}{PE\star X},
\end{equation}
where $PE$ is the Permutation Entropy, and $\star$ denotes a correlation function across $g$. To account for the finite range in $g$, we take repeating boundary conditions; that is, for both factors in the correlation functions, we wrap around the standard $g$-range of $0.25-0.60$, with the offset $dg$ determining where the two functions overlap. Of course, this results in an artificial increase in the auto-correlation/self-overlap as $|dg|$ gets very large, with $K_{dg=0.175}=K_{dg=-0.175}$ and to suppress the false signal from this artificial effect we display $K$ only for $|dg|<0.1$. The important point to note is that at an offset of $dg\approx\pm0.015$ there are local minima in $\Psi\star\Psi$ and $K(\Psi)$, indicating that $\Psi$ disagrees with the PE especially at these values. That is, increases in $\Psi$ correlate to decreases in PE $0.015$ to the left or right in the $g$-range, and thus the behavior of the Symmetry in Word populations is correlated with -- albeit weakly -- nearby chaos-stability shifts. This `dip' at $dg\approx 0.01$ is exactly consistent with our visual observations and discussions above. To the best of our knowledge, this is the first recorded instance of an ability to `peer around the corner' in parameter space to identify chaos-stability transitions.

\section{Comparison with Other Measures}
To understand whether this tantalizing potential ability to `peer around the corner' at the existence of stable behavior at a {\em neighboring} parameter is visible in other more standard measures and diagnostics, we have compared what we see in the Word populations with a variety of some of the most common tools used to distinguish between chaos and regularity. Specifically, we have considered if such behaviors are visible in Poincar\'{e} sections (PS), and Fourier transforms (FT). PS can be of various sorts, but for a driven system it measures the position and momentum of a system stroboscopically at regular intervals determined by the system's natural frequency. It is also typical to use Fourier transforms on observational time-series to characterize the dynamics via its spectral decomposition. While we do not report details, neither of these showed the sort of distinct change in behavior visible in Word populations, and so we searched further in order to understand if variations on the standard diagnostics might better help us understand our results in the previous section. To this end we developed (somewhat) new quantifiers aimed at getting more direct insight into the temporal behavior of interpeak interval, specifically Generalized Poincar\'{e} Sections (GPS) and Interval Trio Maps (ITM). Generalized Poincar\'{e} Sections (GPS), like normal PS represent the intersections of the trajectory with a plane in phase-space; in this case, unusually for a driven system, we record $x[t]$ and $t\mod{\frac{2\pi}{w}}$ values whenever $\pdv{x}{t}$, or $p[t]$ equals zero (as opposed to using a certain phase for the sinusoidal driving for recording $x$ and $p$). This is merely a rotation of perspective on the trajectory: instead of displaying place and speed at regular time, it shows place and time at regular speed. The relevance of shifting to this GPS is that it records time values at peaks and valleys of the oscillator, thus recording in fact the same sort of information as used to construct interpeak Words, and thus arguably liable to show similar behavior.
\begin{figure}
\centering
\includegraphics[width=120pt]{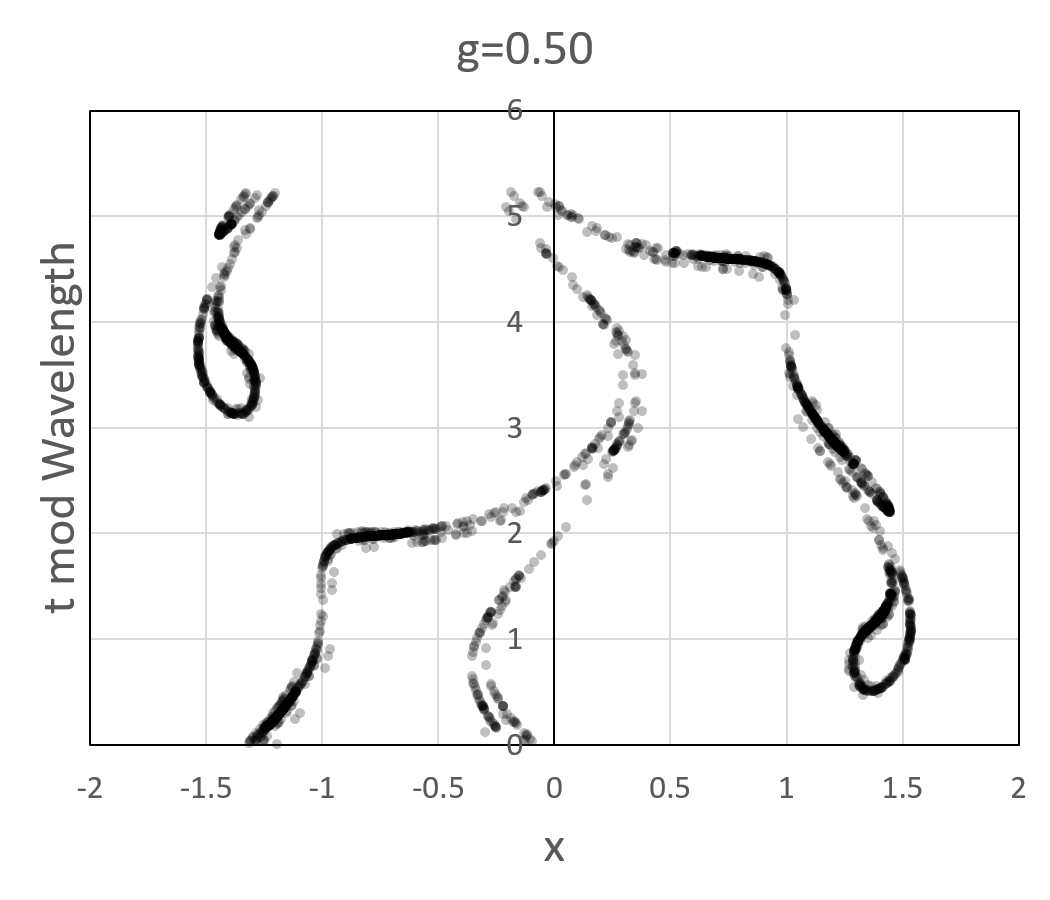}
\includegraphics[width=120pt]{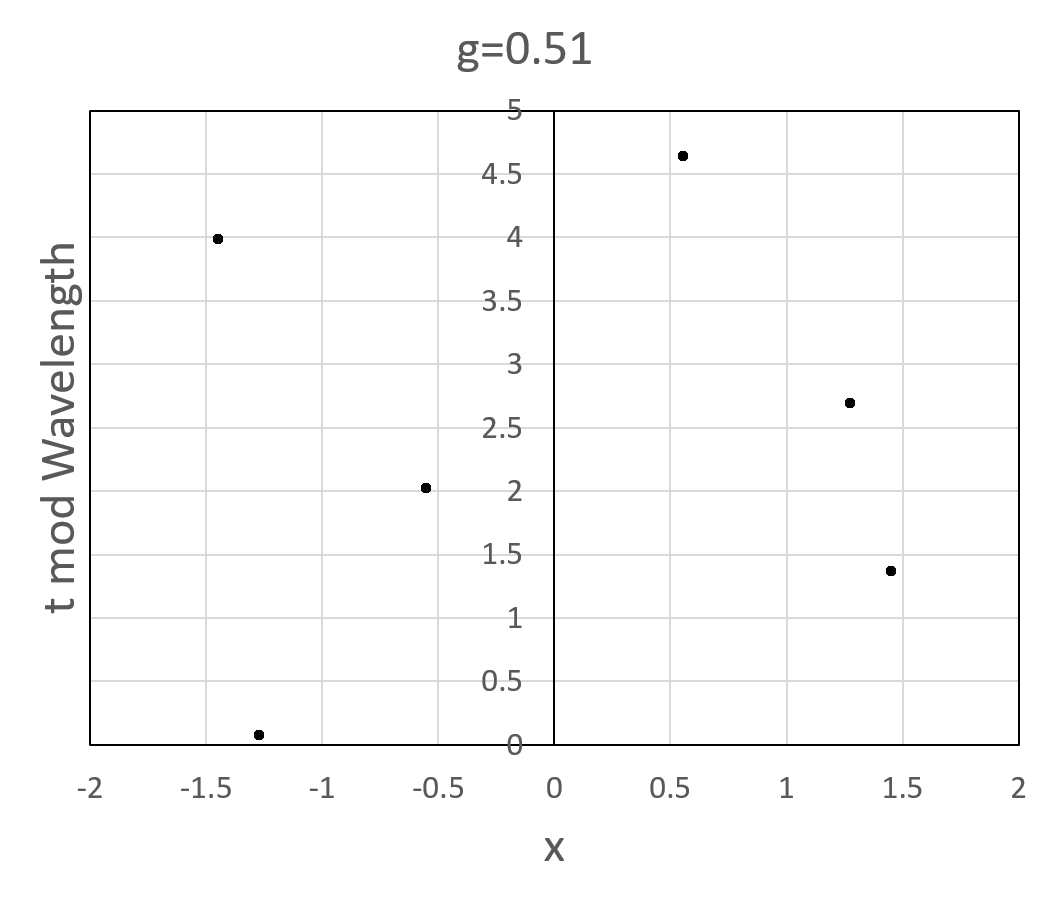}
\caption{The Generalized Poincar\'{e} Sections for $g$-values 0.50 and 0.51. The implied time-delayed symmetry at $g=0.50$ is a result of the fact that GPS takes data at both peaks and valleys in the time-series. Although few points are visible at $g=0.51$, the amount of information depicted is roughly equal to that of $g=0.50$, greater than 1500 data points each. At $g=0.50$, the LE is positive, while at $g=0.51$, it is zero.}
\label{fig:gps}
\end{figure}

Similarly, Interval Trio Maps track interpeak intervals much like Words, but display numerical information, rather than merely ordinal. That is, an ITM is a 3-dimensional representation of trios of consecutive interpeak intervals: every trio is a point in interval-space, where each dimension of that space represents the length of the corresponding interval in the trio. The interval-space of ITMs is divided by three diagonal planes, where two intervals are of equal length: 1st-2nd, 1st-3rd, and 2nd-3rd, equivalent to the planes defined by the equations
\begin{equation}
x=y,x=z,y=z.
\end{equation}
These diagonal planes divide ITMs into six sectors (similar to octants, but only six because the three planes are not orthogonal), with every point in each sector mapping to an instance of an ordinal pattern or Word such that the population associated with a given word amounts to the count of points in a given sector. 
\begin{figure}
\centering
\includegraphics[width=120pt]{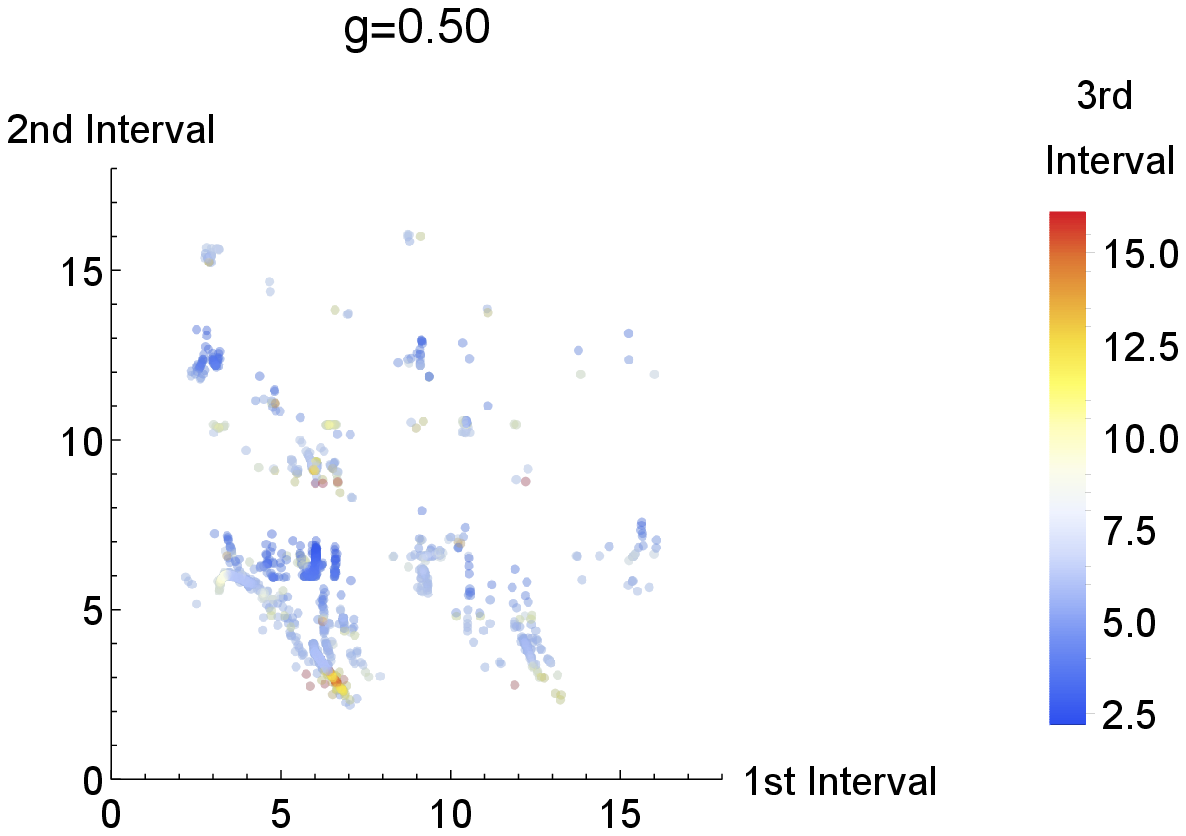}
\includegraphics[width=120pt]{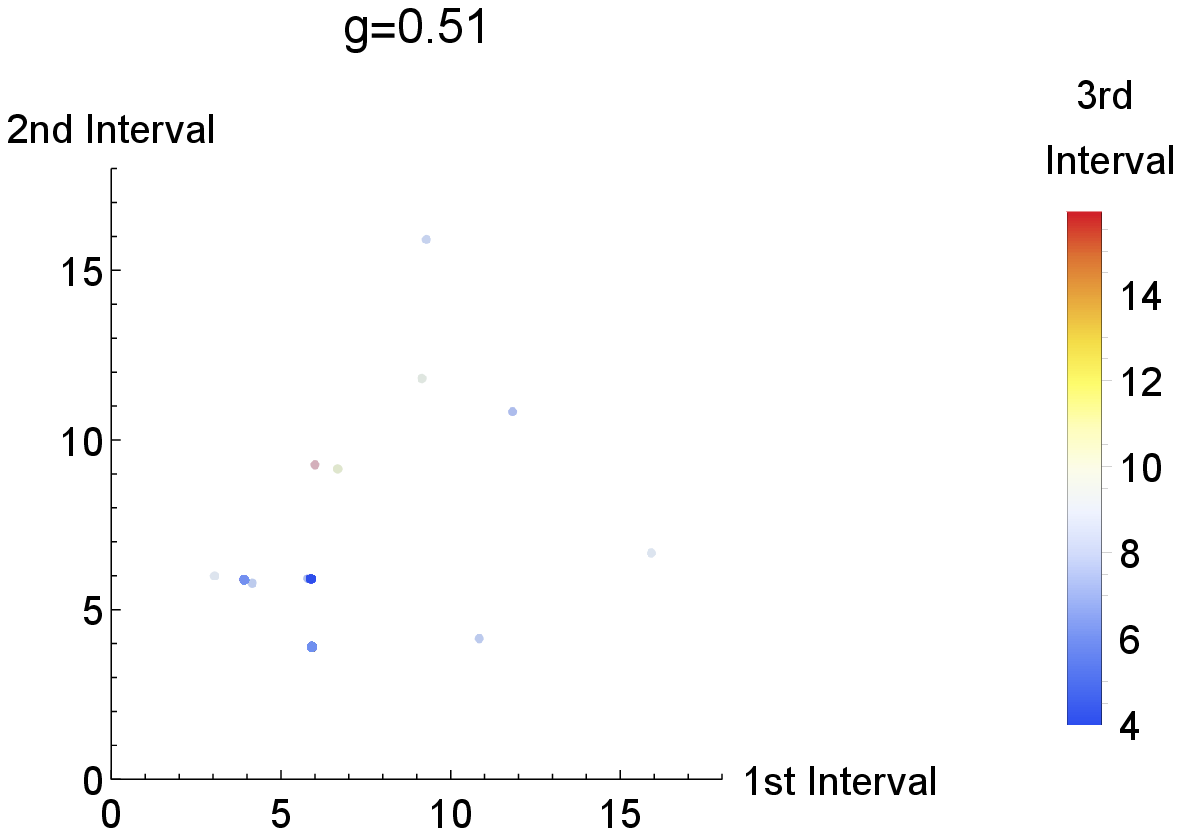}
\caption{The Interval Trio Maps for $g$-values $0.50$ and $0.51$. Warm colors correspond to longer third intervals. At $g=0.50$, the LE is positive, while at $g=0.51$, it is zero.}
\label{fig:itm}
\end{figure}

We considered the same $g$-range as for Word populations for both ITMs and GPSs. We find that indeed ITMs, LEs, and PSs, as well as GPSs, all behave similarly with regard to chaos-stability shifts, in that they show the presence or absence of chaos as a binary -- that is, whether or not there is chaos. This is in contrast to the acute sensitivity visible in Word populations that quantifies greater and lesser degrees of chaos and different types of stability (even at the same LEs). This is significant because it means that Words show a type of structural distinction that is invisible to other metrics. The ITM and GPS both exhibit visual changes that can be described qualitatively as a cloudiness or smearing at the same $g$- and $w$-values as the PS, with small, gradual changes of shape across chaotic zones. PS, because they project a large number of data points across a phase space rather than condensing their information into a single number, contain rather more information, but this information often becomes obscured behind subtle changes, and again, the clearest indicator-- the smearing effect -- only indicates the presence or absence of chaos and perhaps the size and range of the attractor at a given system parameter. This agreement between GPS and PS is expected, if less interesting than if they had disagreed in some way. That ITMs, which contain strictly more information than Words (because Words can be directly derived from ITMs, but not vice versa), do not {\em noticeably} show this sensitivity, may initially be puzzling. However, studying the ITMs allows a little more insight into what we have observed. In particular, smooth changes in the ITM which do not seem noticeably visibly alter it  can sometimes yield drastic changes in Word populations. This is because which Word an Interval Trio belongs to is dependent on its location as a point in interval-space. If an Interval Trio is on or near the boundaries that coincides with all three planes, then all three intervals are of equal length and the Word created is $w_{111}$, but a small deviation from this equality by any two of the three intervals could move the point into one of the six sectors, such that Word population changes amount to changes in higher-order correlations in the statistics of ITMs. This high-order correlation seem to explicate slight regime changes in both stable and chaotic parameter space, whence this acute sensitivity provides -- via the observed correlation -- an advantage in predicting stability in nearby parameter regimes for a system otherwise only understood as being chaotic.

\section{Conclusions and Discussion}
To revisit the original data in the light of all this, there are three major sub-ranges in our $g$ scan that illustrate our findings most efficiently: $0.29-0.31$, $0.33-0.35$, and $0.49-0.53$. At $0.29$, $0.35$, and $0.51-0.53$, the Duffing oscillator is stable, and this is visible in the fact that Word populations are highly stratified and clustered in their frequencies. At $0.35$ for instance, $w_{021}$, $w_{102}$, and $w_{210}$ all have exactly a $\frac{1}{3}$-chance of occurring, and the other three Words are totally forbidden. This is perfect rotational symmetry, and similarly at $0.29$ we see strong mirror symmetry. Then, at $0.30$, $0.34$, and $0.50$, the Duffing oscillator is chaotic as classified by the LE, PE, and the other standard measures. However, we see lingering, lesser forms of the symmetries in Word populations, which are weaker forms of the symmetry extant at neighboring parameters corresponding to stable dynamics. These symmetry groupings reduce in strength and drift closer to the mean as $g$ changes deep into the chaotic regime. For example, at $0.30$, although two of the mirror symmetry groups are still in agreement (although $w_{012}$ and $w_{210}$ are no longer forbidden as they were at $0.29$), the group of $w_{120}$ and $w_{021}$ has dissolved, no longer having similar populations. The fact that the dynamics do not show forbidden patterns can be interpreted as a loss of structure or determinism in the dynamics\cite{Kulp(2016)}. This trend of symmetries, lessening deeper into chaotic zones, continues to $0.48$ and $0.32$, which a deeper than the edge values of the chaotic regime, $0.49-0.50$ and $0.33-0.34$, and which have Word populations that are roughly  $\frac{1}{6}$, compatible with a truly random process. Thus, Word populations, which are constructed by measuring interpeak intervals in a time-series, form distinctive distributions that correlate to chaos and stability, but also distinguish between different types of stable orbits and different levels of temporal structure in chaos. These types of order often come in the form of rotational and mirror symmetries in the Word populations, which can be numerically measured by certain metrics. Furthermore, although Word populations, ITMs, GPSs, and more traditional measures like LE and PS all agree in the places where they detect chaos and stability, no analysis but Word population analysis seems to be able to visibly signal the existence of different kinds of stability and chaos.

Having shown that Word populations and their derived measures are visibly sensitive to regime changes within and between stability and chaos, beyond the blunt binary measures of LE, PS, FT, and ITM, we now consider the relevance of this feature in practical (albeit heretofore hypothetical) applications. In any setting involving a time-series with changing or changeable parameters, where the presence or type of temporal structure is the object of interest, Words seem to prove both simple to construct and useful beyond more traditional measures. Examples would be the ability to provide changed predictive abilities for systems such as precipitation patterns under climate change, $N$-body gravitational problems with non-constant masses, and bridges swaying under varying amounts of foot- and motor-traffic. Whether these systems form synchronous patterns or purely chaotic oscillations depends on both initial conditions and parameter values. Where LEs (in situations where they can be properly measured to begin with) only tell whether any chaos is present, and PSs and ITMs only obliquely distinguish between types of stability and chaos, Word populations display this information clearly and numerically, and can even indicate the proximity of stability. Specifically, the correlation found between Word symmetries in chaos and nearby stability, but with sufficient data fidelity in parameter space, where false signals are rare enough, shows a tangible advantage above the previously mentioned measures. If, for instance, engineers seek to determine whether a bridge could withstand a small increase in foot-traffic without forming resonance and ultimately breaking, an LE or PS, or even an FT, would only say whether the current oscillations are chaotic or not. Words, on the other hand, have Symmetries whose presence correlates with nearby stability, and so could display an increase in symmetries as foot-traffic increases from zero, to the current limit, and towards the proposed new limit, indicating the likelihood of resonance.

This correlation is, to be clear, weak even though it is notably and quantifiably visible across a representative range of parameters. It bears repeating that some false negative and false positive signals show up in the $g$- and $w$-ranges, especially with low data fidelity. To the degree that distinct regions of chaos and stability exist in parameter space, this problem can perhaps be remedied by picking the parameter space with a fine toothed comb, which in our case meant scanning the $g$- and $\omega$-spaces around false signals in increments of $0.001$ rather than $0.01$ steps. However, if a parameter space is densely populated with both chaos and stability -- resembling the black and white distributions of TV static more than zebra stripes -- then the resolution of the parameter scan is irrelevant, as is the concept of "nearness" to stability and chaos. This may be idiosyncratic to the Duffing oscillator, but is generically likely to be true. 

Thus, the limits and conditions for the utility of Words analyses, especially as predictors of chaos-stability transitions remain to be tested more widely. Other than testing other dynamical systems, of course, we also need to consider how the observed correlation between Word symmetries and stability behaves in a higher-dimensional parameter-space, including the possible role of the resolution in parameter space that allows us to pick out these correlations. Other factors which may be need to understand include the role of the minimum resolution for time in the time-series, below which periodic Words grow so populous that hierarchalizing interpeak intervals becomes inefficient. Despite these open questions, it is clear that exploring Word populations in dynamical time series continues to yield a surprising wealth of information about a system's dynamics.

\begin{acknowledgments}
We wish to acknowledge internal support at Carleton College for funding student research. The data that support the findings of this study are available from any of the authors upon reasonable request.
\end{acknowledgments}


\nocite{*}
\begin{thebibliography}{}
\bibitem{Boffetta(2002)}
Boffetta, Guido et al. Predictability: a way to characterize complexity. {\em Physics Reports} {\bf 2002}, {\em 356}(6), 367-474.

\bibitem{Shinbrot(1993)}
Troy Shinbrot, Celso Grebogi, James A. Yorke \& Edward Ott. Using small perturbations to control chaos. {\em Nature} {\bf 1993}, {\em 363}, 411–417.

\bibitem{Hayes(1993)}
Scott Hayes, Celso Grebogi, and Edward Ott. Communicating with chaos. {\em Phys. Rev. Lett.} {\bf 1993}, {\em 70}, 3031.

\bibitem{Cavalcante(2013)}
Hugo L. D. de S. Cavalcante, Marcos Ori\'a, Didier Sornette, Edward Ott, and Daniel J. Gauthier. Predictability and Suppression of Extreme Events in a Chaotic System. {\em Phys. Rev. Lett.} {\bf 2013}, {\em 111}, 198701.

\bibitem{Wang(2016)}
Le-Zhi Wang, Ri-Qi Su, Zi-Gang Huang, Xiao Wang, Wen-Xu Wang, Celso Grebogi and Ying-Cheng La. A geometrical approach to control and controllability of nonlinear dynamical networks. {\em Nature Comm.} {\bf 2016}, {\em 7}, 11323.

\bibitem{Wolf(1985)}
Wolf, Alan et al. Determining Lyapunov exponents from a time series. {\em Physica D Nonlinear Phenomena} {\bf 1985}, {\em 16}(3), 285-317.

\bibitem{Bandt(2002)}
Bandt, Christoph and Bernd, Pompe. Permutation entropy: a natural complexity measure for time series. {\em Phys Rev Lett} {\bf 2002}, {\em 17}, 1--4.

\bibitem{Amigo(2006)}
Amig\'{o}, Jos\'{e} et al. Order patterns and chaos. {\em Physics Letters A} {\bf 2006}, {\em 355}, 27--31.

\bibitem{Amigo(2010)}
Amig\'{o}, Jos\'{e}. Detection of determinism. In {\em Permutation Complexity in Dynamical Systems}; Springer {\bf 2010}: Berlin, Germany; pp. 147--158.

\bibitem{Zou(2019)}
Yong Zou, Reik V. Donner, Norbert Marwan, Jonathan F. Donges, J\"urgen Kurths. Complex network approaches to nonlinear time series analysis. {\em Physics Reports} {\bf 2019}, {\em 787}, 1-97.


\bibitem{Trostel(2018)}
Max L. Trostel, Moses Z. R. Misplon, Andr\'es Aragoneses, and Arjendu K. Pattanayak. Characterizing complex dynamics in the classical and semi-classical duffing oscillator using ordinal patterns analysis. {\em Entropy} {\bf 2018}, {\em 20}, 40.

\bibitem{Barreiro(2011)}
Marcelo Barreiro, Arturo C. Marti, Cristina Masoller. Inferring long memory processes in the climate network via ordinal pattern analysis. {\em Chaos} {\bf 2011}, {\em 21}, 013101.

\bibitem{Soriano(2011)}
Miguel C. Soriano, Luciano Zunino, Osvaldo A. Rosso, Ingo Fischer, and Claudio R. Mirasso. Time Scales of a Chaotic Semiconductor Laser with Optical Feedback Under the Lens of a Permutation Information Analysis. {\em IEEE Jour. Quant. Elec.} {\bf 2011}, {\em 47}, 2, 252.

\bibitem{Parlitz(2012)}
Parlitz, U.; Berg, S.; Luther, S.; Schirdewan, A.; Kurths, J.; Wessel, N. Classifying cardiac biosignals using ordinal pattern statistics and symbolic dynamics. {\em Comput. Biol. Med.} {\bf 2012}, {\em 42}, 319–327.

\bibitem{Tirabassi(2016)}
Giulio Tirabassi and Cristina Masoller. Unravelling the community structure of the climate system by using lags and symbolic time-series analysis. {\em Sci. Rep.} {\bf 2016}, {\em 6}, 29804.

\bibitem{Colet(2018)}
Meritxell Colet and Andr\'es Aragoneses. Forecasting events in the complex dynamics of a semiconductor laser with optical feedback. {\em Sci. Rep.} {\bf 2018}, {\em 8}, 210741.

\bibitem{Guckenheimer(1983)}
J. Guckenheimer and Philip Holmes. Nonlinear Oscillations, Dynamical Systems and Bifurcations of Vector Fields, {\em Springer}, {\bf 1983}

\bibitem{Bandt(2015)}
Christoph Bandt. Permutation entropy and order patterns in long time series. In I. Rojas and H. Pomares,
editors, Time Series Analysis and Forecasting, Contributions to Statistics. Springer, {\bf 2015}.

\bibitem{Bandt(2019)}
Christoph Bandt. Small order patterns in big time series: a practical guide. {\em Entropy} {\bf 2019}, 21(6): 613.

\bibitem{Kulp(2016)}
C. W. Kulp, J. M. Chobot, B. J. Niskala, and C. J. Needhammer Using forbidden ordinal patterns to detect determinism in irregularly sampled time series. In {\em Chaos}; {\bf 26}, 023107 (2016)
\end{thebibliography}
\end{document}